\documentclass[fleqn,usenatbib,useAMS]{mnras}

\usepackage{newtxtext,newtxmath}

\usepackage[T1]{fontenc}

\DeclareRobustCommand{\VAN}[3]{#2}
\let\VANthebibliography\thebibliography
\def\thebibliography{\DeclareRobustCommand{\VAN}[3]{##3}\VANthebibliography}

\usepackage{graphicx}	
\usepackage{amsmath}	
\usepackage{xcolor}
\usepackage{multirow}
\usepackage{multicol}
\usepackage{caption}
\usepackage{subcaption}
\usepackage{url}

\title[Cosmology with future Rubin LSST]{Role of Future SNIa Data from Rubin LSST in Reinvestigating Cosmological Models}

\author[R. Shah, A. Mitra, P. Mukherjee, B. Pal, S. Pal]{
Rahul Shah,$^{1}$\thanks{E-mail: rahul.shah.13.97@gmail.com}
Ayan Mitra,$^{2,3,4}$\thanks{E-mail: ayan.mitra@iucaa.in}
Purba Mukherjee,$^{1}$\thanks{E-mail: purba16@gmail.com}
Barun Pal,$^{5}$\thanks{E-mail: terminatorbarun@gmail.com}
and Supratik Pal$^{1,6}$\thanks{E-mail: supratik@isical.ac.in}
\\
$^{1}$Physics and Applied Mathematics Unit, Indian Statistical Institute, 203 B.T. Road, Kolkata 700 108, India\\
$^{2}$Center for AstroPhysical Surveys, National Center for Supercomputing Applications, University of Illinois Urbana-Champaign, Urbana, IL, 61801, USA\\
$^{3}$Department of Astronomy, University of Illinois at Urbana-Champaign, Urbana, IL 61801, USA\\
$^{4}$Kazakh-British Technical University, 59 Tole bi street, Almaty 050000, Kazakhstan\\
$^{5}$Netaji Nagar College for Women, 170/13/1 N.S.C. Bose Road, Regent Estate, Kolkata-700092, India\\
$^{6}$Technology Innovation Hub on Data Science, Big Data Analytics and Data Curation, Indian Statistical Institute, 203 B.T. Road, Kolkata 700 108, India
}

\date{Accepted 2024 April 10. Received 2024 February 11; in original form 2023 May 20}

\pubyear{2024}

\begin{document}
\label{firstpage}
\pagerange{\pageref{firstpage}--\pageref{lastpage}}
\maketitle

\begin{abstract}
We study how future Type-Ia supernovae (SNIa) standard candles detected by the Vera C. Rubin Observatory (LSST) can constrain some cosmological models. We use a realistic three-year SNIa simulated dataset generated by the LSST Dark Energy Science Collaboration (DESC) Time Domain pipeline, which includes a mix of spectroscopic and photometrically identified candidates. We combine this data with Cosmic Microwave Background (CMB) and Baryon Acoustic Oscillation (BAO) measurements to estimate the dark energy model parameters for two models -- the baseline $\Lambda$CDM and Chevallier-Polarski-Linder (CPL) dark energy parametrization. We compare them with the current constraints obtained from joint analysis of the latest real data from the Pantheon SNIa compilation, CMB from Planck 2018 and BAO. Our analysis finds tighter constraints on the model parameters along with a significant reduction of correlation between $H_0$ and $\sigma_{8,0}$. We find that LSST is expected to significantly improve upon the existing SNIa data in the critical analysis of cosmological models.
\end{abstract}

\begin{keywords}
(cosmology:) cosmological parameters 
-- cosmology: observations
-- (stars:) supernovae: general 
-- (cosmology:) dark energy 
-- methods: data analysis 
-- instrumentation: detectors 
\end{keywords}



\section{Introduction} \label{intro}

Type Ia Supernova (SNIa) data has played a crucial role in advancing our understanding of the Universe's expansion history and the nature of dark energy (DE). \citet{riess_1998} and \citet{perl} published their landmark discovery in which they used SNIa data to measure the deceleration parameter, to establish, for the first time, that the Universe is currently expanding at an accelerated rate. The combined results provided strong evidence for cosmic acceleration, the cause of which is widely considered to be the existence of a mysterious ``dark energy'' component. Following this discovery, there has been an ever-expanding effort to improve the precision of SNIa data and to tighten dark energy constraints. The Union 2.1 \citep{2012ApJ...746...85S} compilation, released in 2011, included data from 580 SNIa and was the largest and most comprehensive sample at the time. This was followed by the Joint Light-curve Analysis (JLA) \citep{Betoule2014}, with a compilation of $740$ spectroscopically confirmed SNIa, designed to have a set of observations with consistent methods of calibration and analysis, which helped improved the precision of SNIa measurements. In 2018, the Pantheon \citep{Pantheon} compilation was released, which included data from 1048 SNIa covering a wide range of redshifts up to $z\sim2.3$, and improved the precision of the measurements further. More recently, the Dark Energy Survey (DES) \citep{DES:2018paw, DES:Y3DR} collaboration published their Y3 cosmology data release in 2019, which included $207$ spectroscopically confirmed SNIa and supplemented with additional $122$ low-$z$ SNIa from external catalogues resulting in DES-SN3YR dataset of $329$ total SNIa. A follow-up analysis to Pantheon, called ``Pantheon+'' \citep{pantheon_new}, was released in 2021, which included additional data from the Dark Energy Survey (DES) and further increased the precision of the measurements. The Pantheon+ dataset included data from 1701 light curves from 1550 distinct SNIa and confirmed the results from the original Pantheon study.

Although abundant observational evidence from diverse sources \citep{Kowalski:2008ez,2007ApJS..170..377S,2006PhRvD..74l3507T,Planck2018,2010MNRAS.401.2148P,Fedeli:2008fh} supports SNIa data collectively presenting a persuasive case for the Universe's current acceleration, the exact mechanism behind this phenomenon necessitating a substantial negative pressure continues to elude understanding. Two widely used approaches have been proposed to address this issue, namely introducing an additional cosmic species to the energy budget of the Universe, called dark energy, and modifying the theory of gravity itself \citep{Starobinsky:1979ty,1970MNRAS.150....1B,Elizalde:2010jx,Arora:2020tuk,Harko:2011kv,Bamba:2010wb,Linder:2010py,Ferraro:2006jd,Nojiri:2003ft}. These approaches have both strengths and limitations, leading to various models attempting to explain the observed late-time acceleration, with new models continuously being proposed. Despite the success of the six-parameter baseline Lambda Cold Dark Matter ($\Lambda$CDM) model in fitting a broad range of data sets, it has faced several limitations, such as the cosmological constant problem \citep{Lombriser:2019jia,Magueijo:2021rpi}, the cosmic coincidence problem \citep{Copeland:2006wr,Velten:2014nra}, the primordial singularity problem \citep{Belinski:2009wj,Craps:2010bg}, and also the growing inconsistencies between measurements of cosmological parameters obtained from different scales of observations \citep{snowmasstensions}. Among these, one of the most notable inconsistencies is the $\sim5\sigma$ tension in the determination of the present value of the Hubble parameter, denoted by $H_0$, where the measurement by the SH0ES team locally gives $H_0=73.3\pm 1.04$ kms$^{-1}$ Mpc$^{-1}$ \citep{Riess_2022}, while the value inferred from the early cosmic microwave background (CMB) radiation sky by the Planck survey yields $H_0=67.36\pm0.54$ kms$^{-1}$ Mpc$^{-1}$ for $\Lambda$CDM \citep{Planck2018}. The precise observations of the Cosmic Microwave Background (CMB) by Planck \citep{Planck2018}, Large Scale Structure (LSS) and Baryon Acoustic Oscillations (BAO) from various surveys \citep{6dF,MGS,BOSS2}, and SNIa data from Pantheon and DES have put moderate level of constraints on the properties of dark energy through joint analyses. Despite its considerable amount of success, we remain far from accurately identifying the precise candidate for dark energy, its equation of state, the necessary parameters for a consistent description of our Universe, and ultimately, the fundamental {\em cosmological model} \textit{per se}. Given the prevalent Hubble tension, as well as other inconsistencies, it is evident that the pursuit of a comprehensive cosmological model is ongoing. An effective cosmological model must demonstrate coherence with all current and forthcoming observations. Consequently, constructing a viable dark energy model presents a significant challenge.

In order to tackle the issues related to vanilla $\Lambda$CDM, models involving dynamical dark energy (DDE) with a varying equation of state (EoS) have been proposed. Several such examples are there in the literature that have drawn varied level of attention from the community \citep{Copeland:2006wr}. Of them, the Chevallier-Polarski-Linder (CPL) parametrization, introduced by Chevallier and Polarski \citep{cpl1} and Linder \citep{cpl3}, has been the most widely used and popular DDE model. In this article we will be focusing on these two models in particular, in order to study their potential in addressing the nature of dark energy using present and future data.

Complementary to the existing data, the next generation of surveys, including Euclid \citep{EuclidTheoryWorkingGroup:2012gxx,https://doi.org/10.48550/arxiv.0912.0914}, LSST \citep{LSSTDarkEnergyScience:2018jkl,Zhan:2017uwu,LSST:2008ijt}, Roman Space Telescope \citep{Akeson:2019biv}, the Thirty Meter Telescope (TMT) \citep{TMT} and the already launched JWST \citep{Gardner_2006}, will provide a 10-fold increase in the precision of dark energy constraints over the Stage 2 surveys \citep{FoM}. The recent discovery of gravitational waves (GW) \citep{2} offers an additional probe for tracing dark energy. The luminosity distance to GW events is directly inferable from the detected signal \citep{Schutz1986} and the redshifts may be obtained either via an electromagnetic counter-detection \citep{Holz:2005df}, or via cross-correlations with surveys of distribution of galaxies \citep{MacLeod:2007jd}. Future GW missions, such as eLISA \citep{eLISA:2013xep,lisa,2017arXiv170200786A}, DECIGO \citep{Kawamura:2020pcg}, Einstein Telescope \citep{Maggiore:2019uih} and Cosmic Explorer \citep{Reitze:2019iox}, promise to detect numerous GW events in the next two decades with increased precisions and at higher redshifts, which would help constrain the expansion history of the Universe, and hence the nature of dark energy. Also, the Square Kilometre Array (SKA) \citep{SKA:2018ckk} will probe the Epoch of Reionisation (EoR) via the 21-cm HI signal, providing another window for probing the expansion history and dark energy. Thus it is important at the current juncture to revisit the popular dark energy models and study their feasibility in light of the upcoming next-generation surveys, combining the power of multiple cosmological probes. 

It is in this context that the upcoming Large Synoptic Survey Telescope (LSST) \citep{LSSTDarkEnergyScience:2018jkl} can play a vital role. Although the role of SNIa data in testing the viability of different cosmological models is more or less well-established for more than two decades now, till date only around $\sim10^3$ SNIa events have been discovered as mentioned earlier. In spite of a certain level of constraining power, the existing SNIa data still leaves room to accommodate a plethora of models as possible dark energy candidates. So, a more rich SNIa dataset can act as a saviour in this context. On top of that, several models have been proposed with an attempt to alleviate the Hubble tension involving existing cosmological datasets. These models must be revisited against future data in order to check their robustness with the advent of data-driven science. LSST - the upcoming SNIa mission - is expected to discover over ten million supernovae during its $10$-year survey program, spanning a broad range in redshift and with uniform photometric calibration, thereby overhauling all its predecessors. This will enable a dramatic step forward in supernova studies, using the power of unprecedented large statistics to control systematic errors and thereby lead to major advances in the precision of supernova cosmology.

Keeping this in mind, we conduct a comprehensive investigation of the performance of two models, $\Lambda$CDM and CPL, in the context of future LSST SNIa data. We begin by obtaining constraints on the model parameters using SNIa from Pantheon, CMB from Planck18, and BAO from various surveys (as elaborated in \ref{datasets}) using {\tt CLASS} \citep{CLASS1,CLASS2} and {\tt MontePython} \citep{MontePython1,MontePython2}. We then explore the impact of the LSST on these models in conjunction with existing data by substituting the Pantheon dataset with LSST and by running the MCMC code afresh. This helps us make a comparison between existing constraints and that expected from LSST. This in turn will reflect the prospects of LSST in revisiting cosmological models under consideration along with their pros and cons. Further, as argued in \citet{Shah:2023}, to avoid any biased analyses, we refrain from incorporating any direct measurement prior on $H_0$ from the SH0ES collaboration's measurement in our parameter estimation, as it can lead to an inherent bias towards higher values of $H_0$, and consequently, on other parameters degenerate with it.

This manuscript is structured in the following manner. Section \ref{data} outlines the LSST data utilised in this study. In Section \ref{models} we outline the cosmological models and the datasets considered in this study. In Section \ref{sec:results}, we derive the constraints on the cosmological models under investigation from the LSST SNIa forecast data, as described in the preceding section, and compare them with the constraints obtained from the Pantheon dataset \citep{Pantheon}. Section \ref{sec:discussion} provides a discussion of our findings and their implications. Finally in Section \ref{sec:conclusion} we present our conclusion.

\section{LSST} \label{data}

\begin{figure*}
    \centering
    \includegraphics[width = 0.6\textwidth]{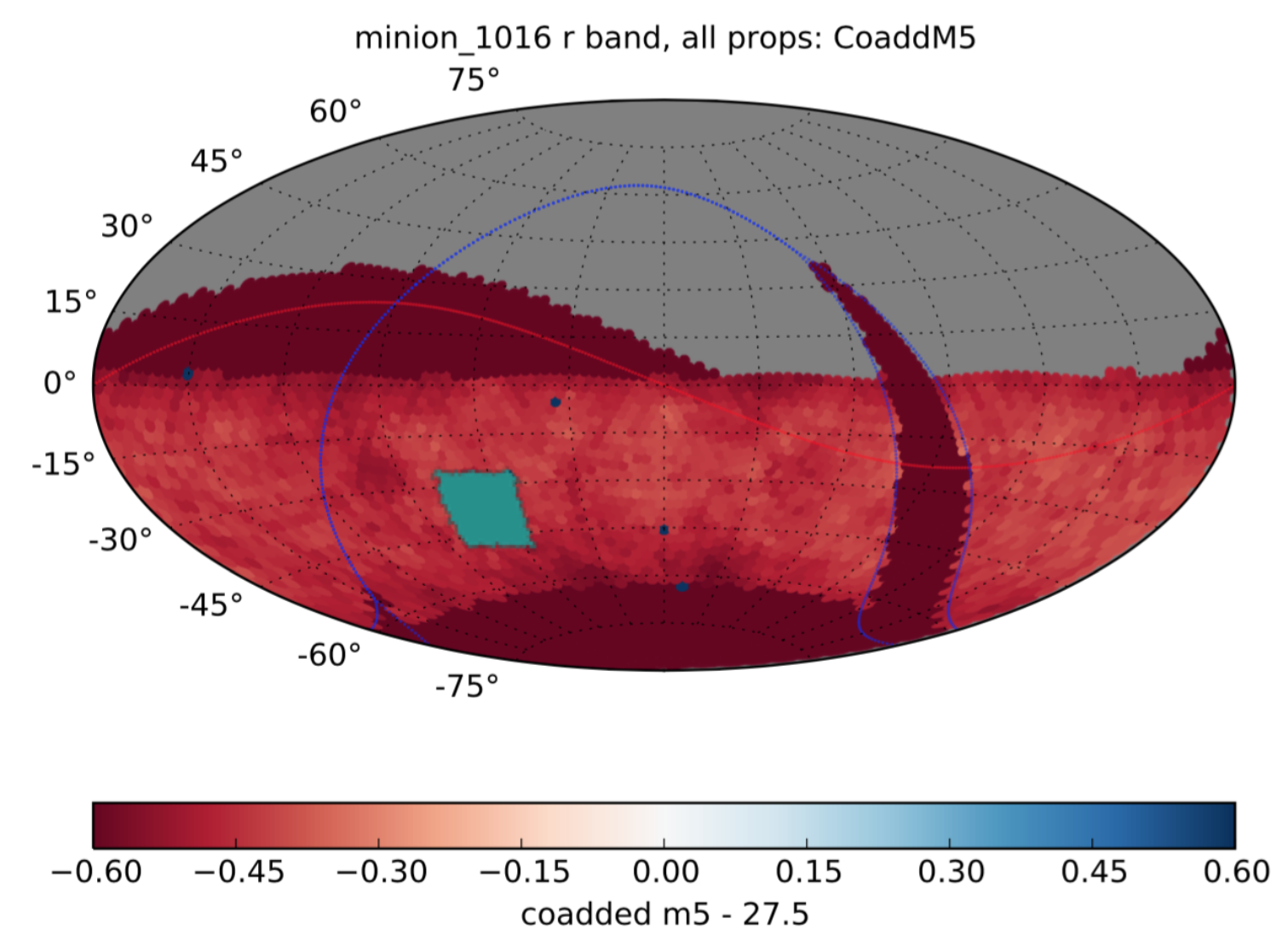}
    \caption{Full sky map showing the potential coverage by LSST observations (in red) from MINION1016\protect\footnote{\url{http://ls.st/Collection-4604}}. The blue and red lines correspond to the Galactic equator and ecliptic respectively. The three small square patches in black each correspond to a 1-degree-square area. There are four such patches in total that will be observed by the LSST; the fourth patch is within the bigger blue patch, which corresponds to the DC2 observations of the LSST. These four small patches make up the Deep Drilling Fields (DDF), which is roughly 50 deg$^2$ in area. Image credit: \citet{LSSTDarkEnergyScience:2020oya}} \label{fig:dc2}
\end{figure*}

\begin{figure*}
    \centering
    \includegraphics[width = 0.6\textwidth]{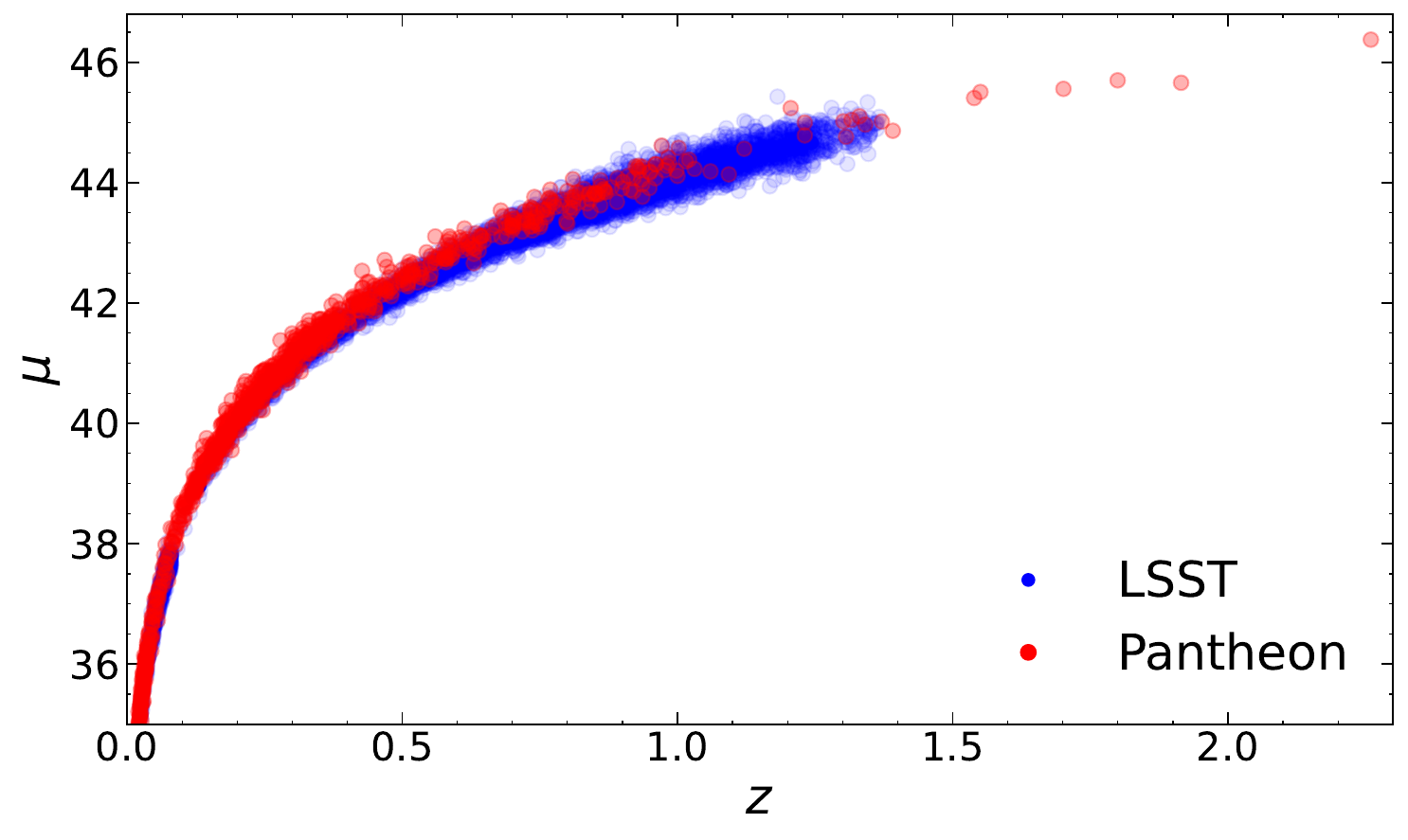}
    \caption{Comparison of the Hubble diagrams between Pantheon (red) and the LSST (blue). The Y-axis corresponds to the distance modulus ($\mu$) plotted as a function of the redshift ($z$) in X-axis.} \label{fig:hd_LSST} 
\end{figure*}

In this analysis, we used simulated $3$-year LSST\footnote{\url{www.lsst.org}} SNIa data to constrain cosmology. The LSST survey is poised to become a defining project of its generation, and it is the most ambitious optical survey currently planned. The survey is planned to begin in 2024 and will undertake a decadal survey. One of the main science themes of the survey is to probe the dark Universe. LSST will be a large, wide-field, ground-based survey that will image the southern sky in six optical passband filters with an unprecedented combination of depth, area, and cadence. During this period, the LSST is forecast to observe millions of supernovae. The Simonyi Survey optical Telescope at the Rubin Observatory includes an $8.4~m$ mirror\footnote{$6.7$~m of effective aperture} and a state-of-the-art $3200$-megapixel camera ($9.6$~deg$^2$ FoV). About $90\%$ of the observing time will be devoted to a deep-wide-fast survey mode that will uniformly observe an $18,000$ deg$^2$ region about $800$ times (summed over all six bands) during the anticipated $10$ years of operations and will yield a co-added map to $r\ - 27.5$. These data will result in databases including about 32 trillion observations of 20 billion galaxies and a similar number of stars, and they will serve the majority of the primary science programs. The remaining $10\%$ of the observing time will be allocated to special projects, such as Very Deep and Very Fast time-domain surveys, whose details are currently under discussion.

\subsection{LSST SN Data: Overview}

This paper utilises simulated SNIa data to forecast dark energy results from LSST observations, sourced from \citet{Mitra:2022ykq}. The SN data is generated using the LSST Dark Energy Science Collaboration (DESC) time domain (TD) pipeline and {\tt SNANA} code \citep{snana}, consisting of four main stages illustrated in Fig. 4 of \citet{Mitra:2022ykq}. For simulations, the following input cosmology is used: $\Omega_m=0.3150, \, \Omega_\Lambda=0.6850, \, w_0=-1, \, w_a=0$. The curvature is computed internally as $\Omega_k=1-\Omega_m-\Omega_\Lambda$. These are based on the cosmological parameter constraints from Planck 2018 \citep{Planck2018}. Additionally, the parameter $H_0$ is set to $70.0~ \text{km s}^{-1} \text{Mpc}^{-1}$, which is tied to SALT2 training. We use the SALT2 lightcurve model and generate the observer frame magnitudes. The noise in the simulation is computed using the following equation, 
\small
\begin{equation}
    \sigma_{\text{SIM}}^2 = \left[ F + (A\cdot b) + (F\cdot \sigma_{\text{ZPT}})^2 + \sigma_0\cdot 10^{0.4\cdot ZPT_{\text{pe}}} + \sigma_{\text{host}}^2 \right]S^2_{\text{SNR}}.
\end{equation}
\normalsize
Here $F$ is the simulated flux in photoelectron (p.e.), $A$ is the noise equivalent area given by $A = \left[2\pi\int \text{PSF}^2 (r,\theta)r \mathop{}\!\mathrm{d} r\right]^{-1}$, where PSF stands for the Point Spread Function, $b$ is the background per unit area (which includes sky + CCD readouts + dark current), $S_{\text{SNR}}$ is en empirically determined scale that depends on the signal-to-noise ratio. The three terms denoted by $\sigma$ correspond to zero point uncertainty, flux calibration uncertainty and underlying host galaxy uncertainty. These quantities are empirically determined through fits that match simulated uncertainties with those derived from the survey designs.

Cosmology fitting, the final stage of the pipeline, was performed by the authors in this analysis. Prior to cosmology fitting, the pipeline includes SN brightness standardisation via a Light Curve (LC) fit stage, simulations for bias correction, and a BEAMS with Bias Corrections (BBC) stage for Hubble diagram production. We used a redshift binned Hubble diagram and the associated covariance matrix produced from the BBC stage to perform cosmological fitting, exploring the two cosmological dark energy models discussed in Sec. \ref{sec:model}.

The SNIa dataset used is composed of spectroscopically ($z_{\rm spec}$) and photometrically ($z_{\rm phot}$) identified SNIa candidates combined together. Based on {\tt PLAsTiCC} \citep{plasticc_announce} ``$z_{\rm spec}$'' sample is composed of two sets of events whose spectroscopic redshifts have an accuracy of $\sigma_{z} \sim 10^{-5}$. The first subset is made up of spectroscopically confirmed events whose redshift is predicted to be accurate by the 4MOST spectrograph \citep{4MOST2}, which is being built by the European Southern Observatory, and is expected to become operational in $2023$. 4MOST is situated at a latitude similar to that of the Rubin Observatory in Chile. The second subset is composed of photometrically identified events with an accurate host galaxy redshift determined by 4MOST. The second subset has about $\sim60$\% more events than the first subset. For the photometric sample, host galaxy photo-$z$ was used as a prior (adapted from \citet{Kessler2010}). The photometric redshift and rms uncertainty was based on \citet{Graham2018_photoz}. The whole simulation was re-done based on the {\tt PLAsTiCC} DDF data\footnote{The LSST has different observing strategies, the deep field and the wide field, called as the DDF (Deep drilling field) and the WFD (Wide Fast Deep) respectively.}. Additional low redshift spectroscopic data was used from the DC2 analysis (simulated with WFD cadence). \citet{Mitra:2022ykq} simulated only SNIa and no contamination (\textit{e.g.} core collapse, peculiar SNe, \textit{etc}) was considered. The covariance matrix used corresponds to the stat+syst where syst is all individual systematics combined. Seven systematics in total were considered, which are detailed in Table 3 of \citet{Mitra:2022ykq}. The HD is binned into $14$ redshift bins and contains a total of $5809$ SNIa candidates, as shown in Fig. \ref{fig:hd_LSST}, composed of both spectroscopic and photometric candidates. 

It is worth mentioning that in their work, \citet{Mitra:2022ykq} introduce a data set with characteristics similar to those outlined in the LSST science roadmap for 1 and 10 years of SNIa cosmology analysis, as described by \citet{LSSTDarkEnergyScience:2018jkl}. However, there are notable differences. While the analysis presented by \citet{LSSTDarkEnergyScience:2018jkl} relies solely on SNIa with spectroscopically confirmed host redshifts, \citet{Mitra:2022ykq} expands the scope by including a comprehensive end-to-end analysis that incorporates both spectroscopic and photometric redshifts of host galaxies. This inclusion of SNIa candidates with host photometric redshifts marks \citet{Mitra:2022ykq} as the first of its kind to provide a complete cosmological analysis with this dual approach.

\section{Cosmological Models and Datasets} \label{models}

In this section, we briefly discuss the two cosmological models that are considered in this analysis. In \citet{Mitra:2022ykq} the authors verified, using Fisher analysis, the validity of the simulated LSST dataset used here for the ensuing models. Using the Pantheon and simulated LSST data separately (along with the other mentioned datasets), we will investigate the constraints on the model parameters.

\subsection{Models} \label{sec:model}

\subsubsection*{\textbf{$\Lambda$ Cold Dark Matter ($\Lambda$CDM)} \label{LCDM}}
We include $\Lambda$CDM as the benchmark model in our study, as we are interested in comparing the performances of the alternative models against the baseline model. Often referred to as the ``concordance model'', it is characterised by a constant EoS for dark energy $w=-1$. The Hubble parameter evolves with redshift as
\begin{equation}
    H^2(z)=H_0^2\left[\Omega_{m0}(1+z)^3+(1-\Omega_{m0})\right]. 
\end{equation}

\subsubsection*{\textbf{Chevallier Polarski Linder (CPL)} \label{CPL}}
The CPL parametrization \citep{cpl1,cpl3}, also known as the $w_0w_a$CDM parametrization, is perhaps the most widely used model after $\Lambda$CDM. It is a two-parameter extension to $\Lambda$CDM with a redshift-dependent DE EoS given by
\begin{equation}
    w(z)=w_0+w_a\frac{z}{1+z}\:\:.
\end{equation}
It can be interpreted as the first-order Taylor expansion of a more generic DE EoS $w(z)$ in terms of the scale factor $a=(1+z)^{-1}$ \citep{Zhai:2017vvt}, that is well-behaved at both high and low redshifts. The background history for this model is parametrized as,
\begin{equation}
\begin{split}    
    H^2&(z)=H_0^2\times \\ &\left[\Omega_{m0}(1+z)^3+(1-\Omega_{m0})(1+z)^{3(1+w_0+w_a)}\exp\left(- \frac{3 w_a z}{1+z}\right)\right].
\end{split}
\end{equation}

\subsection{Observational datasets} \label{datasets}

\begin{table}
        \begin{center}
            {\renewcommand{\arraystretch}{1.5} \setlength{\tabcolsep}{30 pt} \centering
            \begin{tabular}{|c|c|}
                \hline
                \textbf{Parameter}             & \textbf{Prior} \\
                \hline
                $\Omega_{\rm b} h^2$           & $[0.005,0.1]$ \\
                $\Omega_{\rm c} h^2$           & $[0.01, 0.99]$ \\
                $100\theta_{s}$                & $[0.5,10]$ \\
                $\ln\left(10^{10}A_{s}\right)$ & $[1,4]$ \\
                $n_{s}$                        & $[0.5, 1.5]$ \\
                $\tau$                         & $[0.,0.9]$ \\
                $w_0$                          & $[-2, 1]$ \\
                $w_a$                          & $[-3, 3]$ \\
                \hline
            \end{tabular}
            }
            \caption{Uniform priors on the model parameters of $\Lambda$CDM and CPL.} \label{tab:priors}
        \end{center}
\end{table}

The following datasets have been taken into account in the present analysis:

\begin{itemize}
    \item \textbf{CMB:} Cosmic Microwave Background temperature and polarisation angular power spectra, and CMB lensing of Planck 2018 TTTEEE+low l+low E+lensing \citep{Pl18V,Pl18VI,Pl18VIII}.
    \item \textbf{BAO:} Baryon Acoustic Oscillations measurements by galaxy surveys \textit{viz.} 6dFGS \citep{6dF} , SDSS MGS \citep{MGS}, and BOSS DR12 \citep{BOSS1,BOSS2}.
    \item \textbf{Pantheon:} Luminosity distance data of 1048 type Ia supernovae from the Pantheon \citep{Pantheon} catalogue. 
    \item \textbf{LSST:} The realistically simulated distance modulus catalogue from Vera C. Rubin Observatory and Legacy Survey of Space and Time, as discussed in Sec. \ref{data}. The simulated data sample consists of 5809 type Ia supernovae candidates.
\end{itemize}

\section{MCMC Constraints and Results} \label{sec:results}

We obtain the constraints on the model parameters using two dataset combinations, namely Pantheon+CMB+BAO (hereafter referred to as PCB) and LSST+CMB+BAO (hereafter referred to as LCB). The priors considered for the MCMC analyses using {\tt CLASS}\footnote{\url{https://github.com/lesgourg/class_public}} \citep{CLASS1,CLASS2} and {\tt MontePython}\footnote{\url{https://github.com/brinckmann/montepython_public}} \citep{MontePython1,MontePython2} are given in Table \ref{tab:priors}. The corresponding contour plots were generated using {\tt GetDist}\footnote{\url{https://github.com/cmbant/getdist}} \citep{Lewis:2019xzd}.

\begin{table}
\begin{center}    
    \resizebox{0.5\textwidth}{!}{\renewcommand{\arraystretch}{1.55} \setlength{\tabcolsep}{20 pt} 
    \begin{tabular}{c c c c c}
        \hline\hline
        \textbf{Models} & $\Lambda$CDM & CPL \\ \hline
        $\Omega_b h^2$ &  $0.02243_{-0.00014}^{+0.00014}$ & $0.02238_{-0.00015}^{+0.00013}$ \\
        $\Omega_c h^2$ &  $0.1192_{-0.00094}^{+0.00093}$ &  $0.12_{-0.0011}^{+0.0011}$  \\
        $100\theta_{s}$ & $1.042_{-0.00029}^{+0.00029}$  &  $1.042_{-0.00030}^{+0.00029}$\\ 
        $\ln\left(10^{10}A_{s}\right)$ &  $3.048_{-0.014}^{+0.014}$ & $3.043_{-0.015}^{+0.014}$  \\
        $n_{s}$ &  $0.9672_{-0.0038}^{+0.0037}$ &  $0.9653_{-0.0039}^{+0.0039}$  \\
        $\tau$ &  $0.05682_{-0.0075}^{+0.0069}$  &  $0.05356_{-0.0073}^{+0.0071}$  \\
        $w_0$ & -  & $-0.9573_{-0.083}^{+0.075}$ \\
        $w_a$ &  - &  $-0.2862_{-0.27}^{+0.31}$ \\
        \hline 
        $H_0$ &  $67.74_{-0.42}^{+0.42}$ &  $68.32_{-0.83}^{+0.83}$  \\
        $\Omega_{m0}$ & $0.3102_{-0.0056}^{+0.0056}$  &  $0.3065_{-0.0082}^{+0.0078}$ \\
        $\sigma_{8,0}$ &  $0.8104_{-0.0061}^{+0.0060}$ &  $0.8205_{-0.011}^{+0.011}$ \\
        \hline 
        $\chi^2_{min}$ & 3811  &  3811  \\
        $-\ln\mathcal{L}_{min}$ & 1905.5  & 1905.52  \\
        \hline
        \hline
    \end{tabular}
    }
\caption{MCMC constraints on the model parameters of $\Lambda$CDM and CPL using combined Pantheon+CMB+BAO (referred to as PCB) datasets.} \label{tab:pcb_constraints}
    \resizebox{0.5\textwidth}{!}{\renewcommand{\arraystretch}{1.55} \setlength{\tabcolsep}{20 pt} 
    \begin{tabular}{c c c c c}
        \hline\hline
        \textbf{Models} & $\Lambda$CDM & CPL \\ \hline
        $\Omega_b h^2$ &  $0.02245_{-0.00013}^{+0.00013}$ & $0.02238_{-0.00015}^{+0.00014}$  \\
        $\Omega_c h^2$ & $0.1191_{-0.00074}^{+0.00074}$  &  $0.1199_{-0.0010}^{+0.0011}$  \\
        $100\theta_{s}$ & $1.042_{-0.00028}^{+0.00028}$  &  $1.042_{-0.0003}^{+0.0003}$  \\ 
        $\ln\left(10^{10}A_{s}\right)$ & $3.05_{-0.014}^{+0.014}$  &  $3.043_{-0.014}^{+0.014}$  \\
        $n_{s}$ & $0.9676_{-0.0035}^{+0.0035}$  &  $0.9654_{-0.004}^{+0.004}$  \\
        $\tau$ &  $0.05722_{-0.0076}^{+0.0069}$  &  $0.05345_{-0.0074}^{+0.0073}$ \\
        $w_0$ & -  &  $-0.9315_{-0.055}^{+0.054}$\\
        $w_a$ &  - &  $-0.3379_{-0.22}^{+0.23}$  \\
        \hline 
        $H_0$ &  $67.84_{-0.34}^{+0.33}$ &  $67.99_{-0.39}^{+0.39}$ \\
        $\Omega_{m0}$ & $0.3087_{-0.0044}^{+0.0044}$  &  $0.3094_{-0.0044}^{+0.0046}$  \\
        $\sigma_{8,0}$ &  $0.8101_{-0.0060}^{+0.0059}$ &  $0.8175_{-0.0083}^{+0.0086}$  \\
        \hline
        $\chi^2_{min}$ &  2796 &  2796  \\
        $-\ln\mathcal{L}_{min}$  & 1397.74  & 1397.9  \\
        \hline 
        \hline
    \end{tabular}
    }
\caption{MCMC constraints on the model parameters of $\Lambda$CDM and CPL using combined LSST+CMB+BAO (referred to as LCB) datasets.} \label{tab:lcb_constraints}
\end{center}
\end{table}

\begin{figure*}
    \centering
    \hspace{-0.4cm}
    \includegraphics[width = 1.02\textwidth]{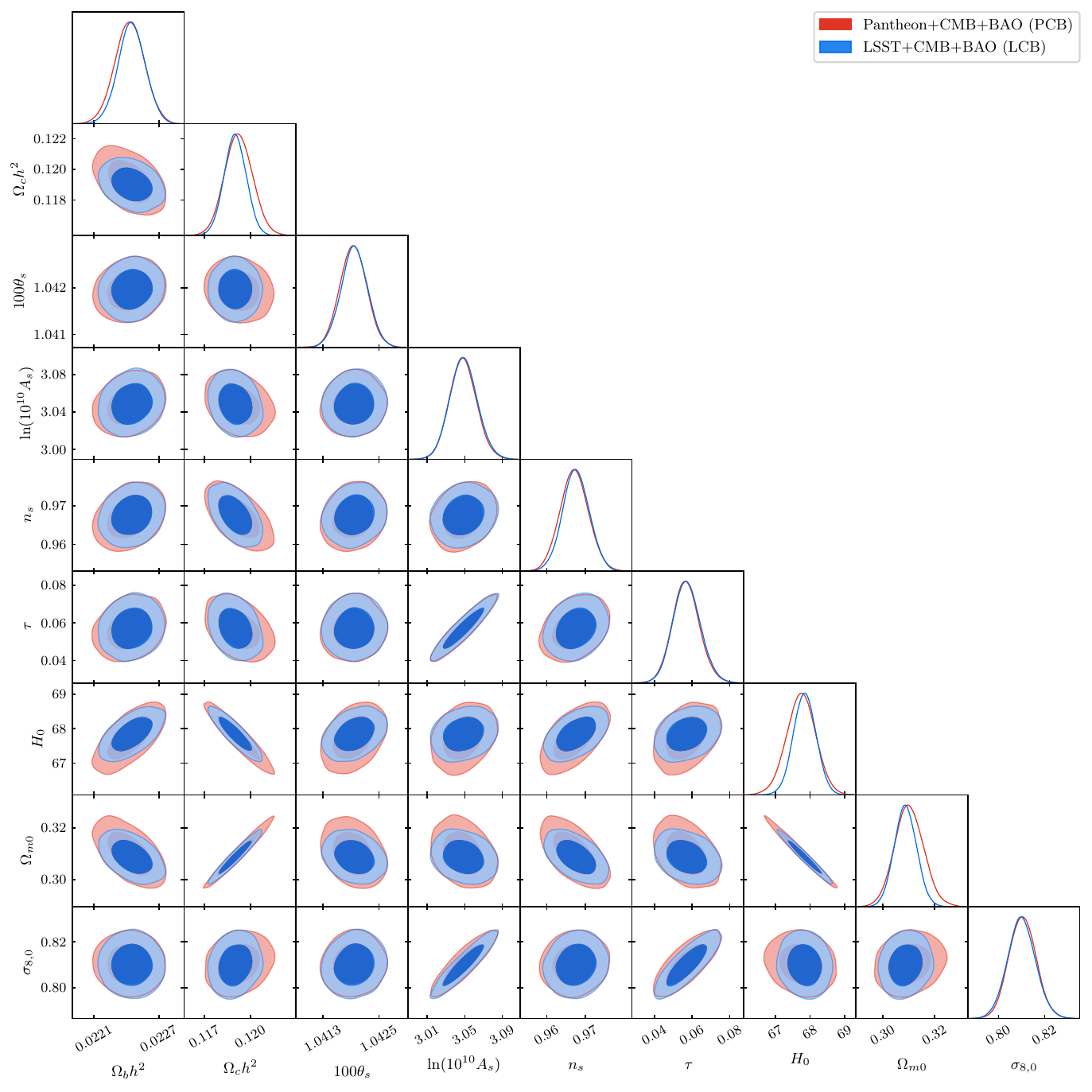}
    \caption{Plots showing the 1-dimensional marginalised posterior distributions and 2-dimensional contour plots for $\Lambda$CDM model parameters using PCB and LCB data sets.} \label{fig:pcb_lcb_lcdm}
\end{figure*}

\begin{figure*}
    \centering
    \hspace{-0.4cm}
    \includegraphics[width = 1.02\textwidth]{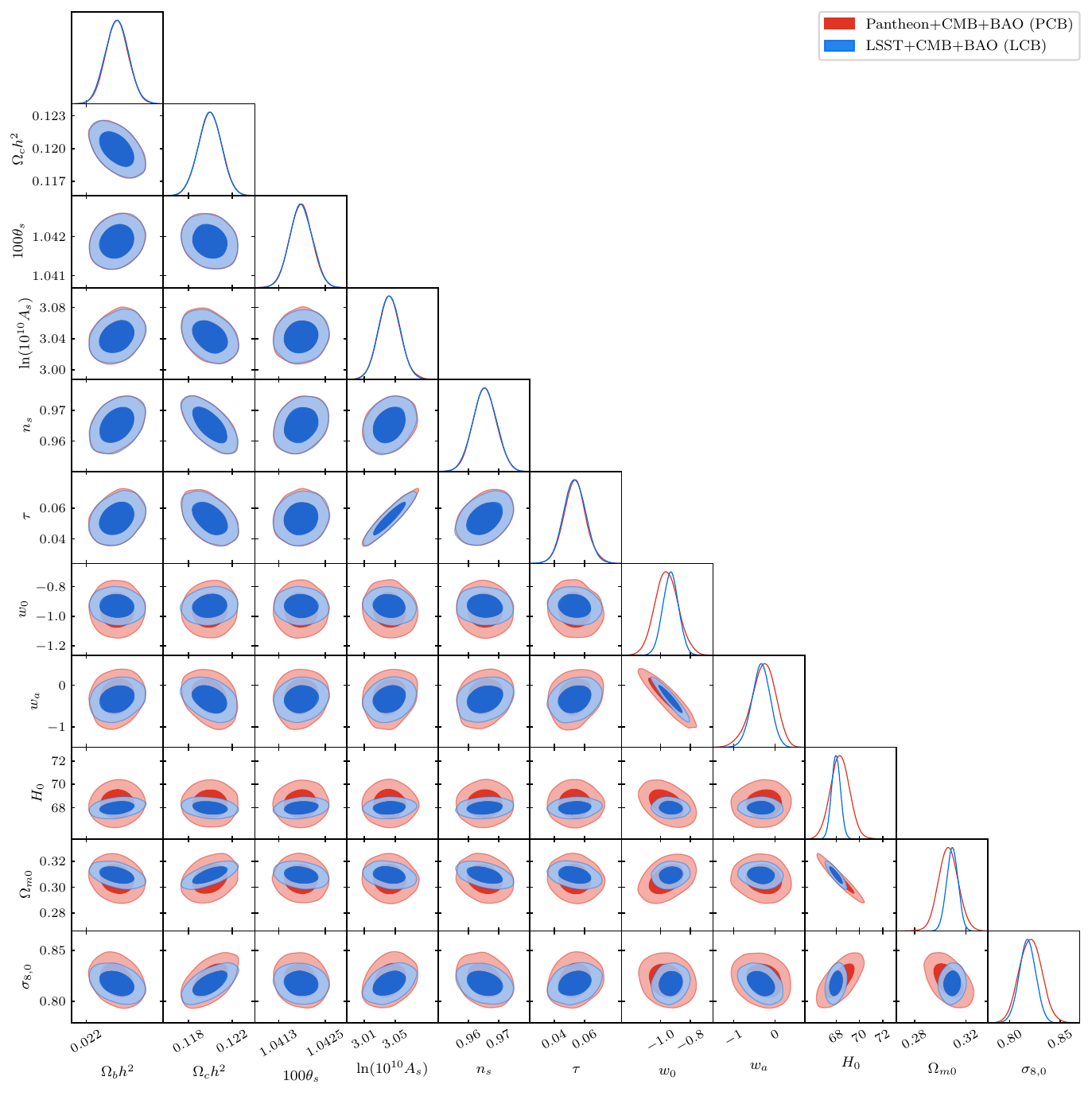}
    \caption{Plots showing the 1-dimensional marginalised posterior distributions and 2-dimensional contour plots for CPL model parameters using PCB and LCB data sets.} \label{fig:pcb_lcb_cpl}
\end{figure*}

Table \ref{tab:pcb_constraints} presents the parameter estimates of the models analysed with the Pantheon dataset (PCB). To ensure equal treatment of the models, we conduct Markov Chain Monte Carlo (MCMC) analyses under identical conditions, enabling the assessment of the effect of our simulated future LSST data when replacing Pantheon. The resulting parameter estimates for the joint LSST+CMB+BAO dataset (LCB) are reported in Table \ref{tab:lcb_constraints}. The corresponding goodness-of-fit parameters, namely the minimum chi-square ($\chi^2_{min}$) and minimum of the negative log-likelihood ($-\ln\mathcal{L}_{min}$) estimates are indicated in their respective tables. Notably, LCB provides a better fit for both models than PCB, as indicated by the goodness-of-fit parameters in the respective tables.

Figure \ref{fig:pcb_lcb_lcdm} displays the $1\sigma$ and $2\sigma$ confidence contours for the full parameter space of $\Lambda$CDM, evaluated with the PCB (in red) and LCB (in blue) datasets. The results reveal a marked improvement in the precision of the cold dark matter energy density $\Omega_c h^2$. This effect is reflected, particularly in the improvement of precision in the derived background quantities, $H_0$ and $\Omega_{m0}$. Nevertheless, their mean values exhibit no significant shifts. No significant variations were observed in the perturbation and reionisation parameters, whether primordial or late-time, \textit{e.g.} $\ln{\left( 10^{10} A_s\right)}$, $\tau$ and $\sigma_{8,0}$. Further, we also observe a slight decrease in the correlation between $H_0$ and $\sigma_{8,0}$ and hence between $\Omega_{m0}$ and $\sigma_{8,0}$ for the LSST dataset in comparison to Pantheon dataset.

\begin{figure*}
\centering
    \includegraphics[width=0.245\textwidth]{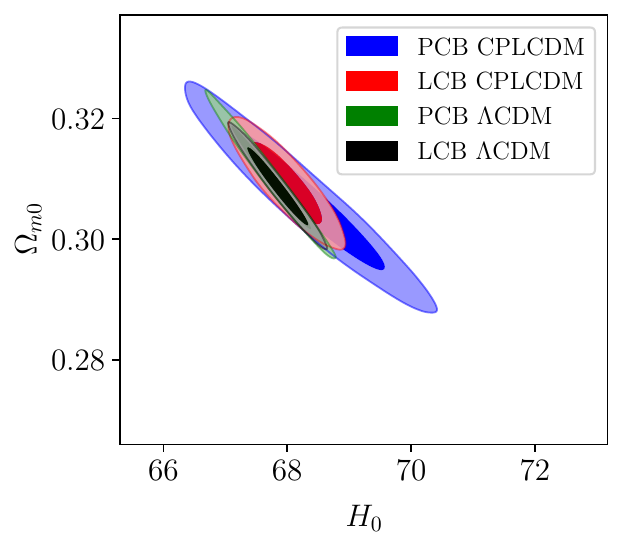}
    \includegraphics[width=0.245\textwidth]{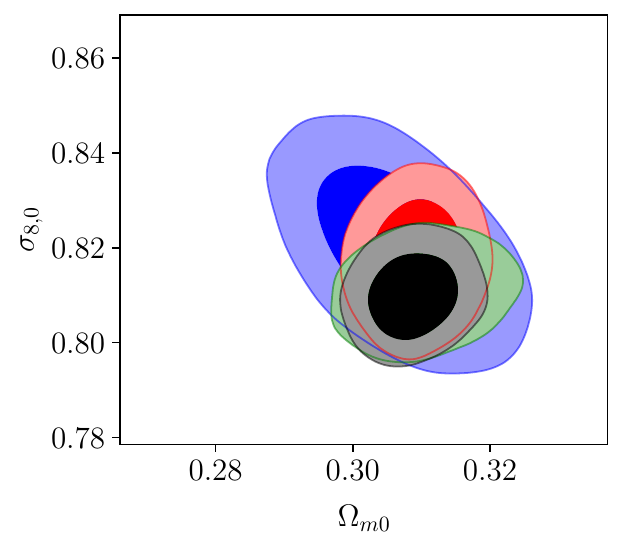} 
    \includegraphics[width=0.245\textwidth]{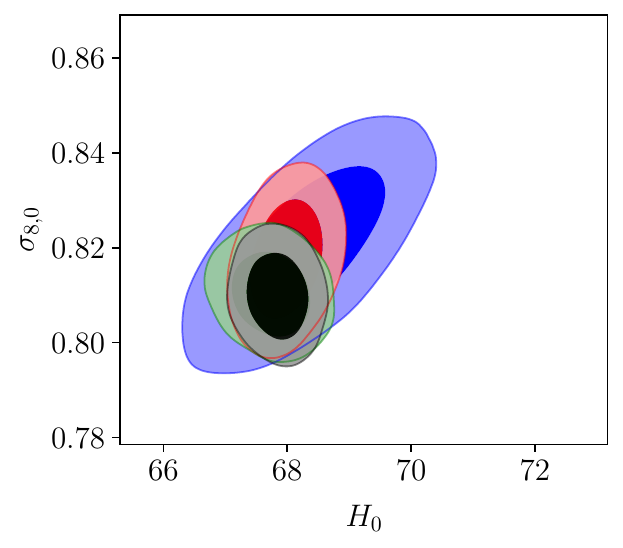}
    \includegraphics[width=0.245\textwidth]{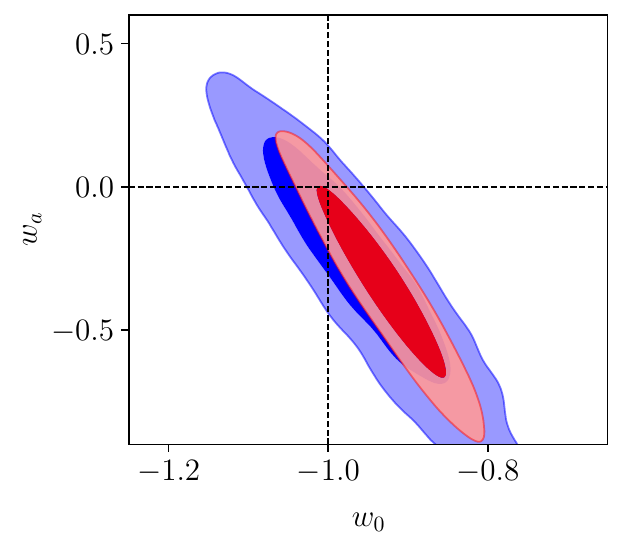}
\caption{Contour plots showing the comparison among the four cases studied.} \label{fig:compare_plot}
\end{figure*}

Figure \ref{fig:pcb_lcb_cpl} is similar to Figure \ref{fig:pcb_lcb_lcdm} but with the CPL dark energy model. We find that except for the dark energy characteristic parameters of the CPL model ($w_0$ and $w_a$) the remaining 6-parameter space exhibits no significant variation with the LCB dataset in comparison to PCB. When compared with baseline $\Lambda$CDM for the same datasets, it hints towards the higher sensitivity of LCB data towards EoS parameters than $\Omega_c h^2$ and that the inclusion of the simulated LSST slightly favours a variable EoS for dark energy. However, since there is no noticeable improvement in $\chi^2$ for CPL compared to $\Lambda$CDM, we refrain from making any strong comments at this point. The parameter space of $w_0$ and $w_a$ becomes strongly restricted when analysed using the LCB combination, with their mean values shifting slightly to almost compensate for each other, with $w_0$ increasing while $w_a$ decreasing. This in turn makes its effect felt on the derived parameters, \textit{viz.} $H_0$, $\Omega_{m0}$ and $\sigma_{8,0}$. It deserves mention that the joint PCB constraints on $H_0$ and $\Omega_{m0}$ for the CPL parametrization are not as tightly constrained as in the case of $\Lambda$CDM; however, the constraints become noticeably more precise when exposed to the combined LCB dataset, with the resulting parameter values being comparable to those obtained for the $\Lambda$CDM scenario. Besides, our results also indicate an increase in precision, along with a slight decrease in the mean value, of $\sigma_{8,0}$ for the LCB dataset compared to PCB. We further notice that the $H_0$-$\sigma_{8,0}$ and $\Omega_{m0}$-$\sigma_{8,0}$ correlations show significant decrease. This feature is intriguing, where the simulated data of LSST is seen to break such correlations to a somewhat considerable extent. This opens up the prospects of LSST in helping re-examine the status of existing alternative cosmological models which are better-performing in the context of $H_0$ and $\sigma_{8,0}$ tensions.
\\\\
Consequently, in Figure \ref{fig:compare_plot}, we show a comparison between the derived parameters $H_0$, $\sigma_{8,0}$ and $\Omega_{m0}$ to parse the correlations and shifts that are impacted by the different analysis choices. We also plot the confidence contours for the CPL parameters, $w_0$ and $w_a$, separately and compare them with the $\Lambda$CDM case ($w_0 = -1$ and $w_a=0$) using dotted lines.

\section{Analysis and Discussion} \label{sec:discussion}

\begin{figure}
    \centering
    \includegraphics[width = 0.485\textwidth]{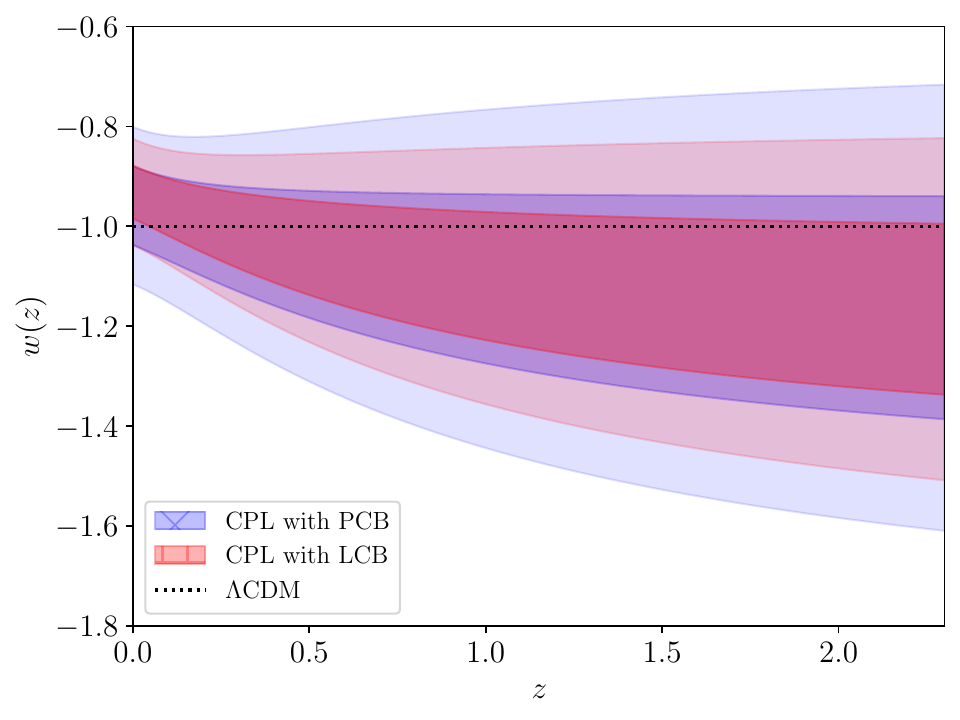}
    \caption{The evolution of the dark energy equation of state $w(z)$ for the CPL parametrizations, taking the values of the model parameters from the MCMC analysis. The shaded regions in red colour with $+$ hatches and in blue colour with $\times$ hatches represent the $1\sigma$ (darker shades) and $2\sigma$ (lighter shades) confidence level for the evolution of $w(z)$ using the combined LCB and PCB data sets. As a reference, we plot the evolution history for $\Lambda$CDM with a dotted line.} \label{fig:wz_plot}
\end{figure}

SNIa data from LSST has the potential to provide a significant improvement over Pantheon, our current standard SNIa dataset, by increasing the number of observed SNIa events and with higher number density of events at high redshifts (up to $z\sim 1.2$ as shown in Fig \ref{fig:hd_LSST}). Furthermore, given the superior instrumentation, we expect the data to be less prone to errors. It is, therefore, crucial to investigate the impact that the future LSST data could have on our understanding of not only the baseline cosmological model but also other popular models that could potentially compete with $\Lambda$CDM and help address the tensions and anomalies associated with it.

In the case of the $\Lambda$CDM model, future LSST data would likely increase the precision of the existing constraints slightly. However, for the CPL model, the impact of the same on precision would be particularly notable. It would result in a stronger constraint on the background, as well as reduced correlations between certain background parameters and perturbation parameters. Since LSST is realistically simulated, we expect these conclusions to remain the same, even for somewhat different fiducial inputs.

Of particular interest is the almost vanishing correlation between $H_0$ and $\sigma_{8,0}$. The existence of a strong positive $H_0$ - $\sigma_{8,0}$ correlation was initially studied by \citet{Bhattacharyya:2018fwb} for the $\Lambda$CDM and CPLCDM models. The presence of this positive correlation implies the intricacy of resolving the Hubble and clustering tensions simultaneously. Their analysis was based on the combination of Planck 2015 data, galaxy BAO measurements and the SNIa data from the JLA compilation. Now, the correlation between two cosmological parameters obtained from an MCMC analysis can be resolved through two distinct approaches. Firstly, by augmenting the dataset with additional information or by enhancing the existing dataset through the incorporation of new data from future missions. This method also tends to narrow down the parameter space, and hence the 1-$\sigma$ range for individual parameters. Alternatively, this correlation can also be addressed by introducing additional parameters and increasing the degrees of freedom in the cosmological model \textit{iff} the datasets under consideration can constrain the same. 

In our work when the $\Lambda$CDM model is constrained with the PCB compilation, we can see that the positive correlation between $H_0$ and $\sigma_{8,0}$, as reported in \citet{Bhattacharyya:2018fwb}, is almost nullified on employing the first approach. With the LCB compilation, we notice a further reduction of this $H_0$-$\sigma_{8,0}$ correlation. For the CPLCDM model, the inclusion of two additional parameters, $w_0$ and $w_a$, drags the matter density parameter $\Omega_{m0}h^2$. This in turn has the effect of shifting the values of $H_0$ and $\sigma_{8,0}$ owing to their existing geometric degeneracy with $\Omega_{m0}$. It is seen that the PCB compilation is unable to nullify the strong positive correlation between $H_0$ and $\sigma_{8,0}$ when compared to \citet{Bhattacharyya:2018fwb}. So, the question of correlations or lifting of degeneracies strongly depends on the cosmological model under consideration. This reveals that the existing latest data sets are unable to address the problems of rising anomalies and tensions in cosmology, through extra modelling of the dark energy EoS. 

However, our analysis with the LCB compilation shows a significant reduction of this correlation in comparison to the PCB case. Although this is an optimistic scenario taking into account simulated data, it appears that the LSST SNIa data would be able to break the existing $H_0$-$\sigma_{8,0}$ (and hence $\Omega_{m0}$-$\sigma_{8,0}$) correlation. This calls for a re-examination of the prospects of existing better-performing cosmological models in the context of the $H_0$ and the $\sigma_{8,0}$ tensions in light of LSST. 

In Figure \ref{fig:wz_plot}, we present the relative performances of the DE EoS $w(z)$for the CPL model, taking the values of the model parameters from the MCMC analysis. The shaded regions in red colour with $+$ hatches and in blue colour with $\times$ hatches show the evolution of $w(z)$ using the combined LCB and PCB data sets, along with the associated uncertainties. The darker shaded regions represent the 1$\sigma$ confidence level, whereas those with lighter shades denote the 2$\sigma$ confidence level. For comparison, we also plot the EoS for $\Lambda$CDM, $w=-1$ with dotted lines. We find significant improvement in terms of precision at both 1$\sigma$ and 2$\sigma$ levels for the LCB over the PCB combination. The DE EoS for the CPL model demonstrates a quintessence nature ($w>-1$) today that tends to cross the phantom divider at higher redshifts. The reconstructed 1$\sigma$ region indicates that the simulated LSST data has a higher preference for a non-phantom EoS at the present epoch with reduced error bars. We observe that the EoS for CPL excludes $\Lambda$CDM ($w=-1$) at 1$\sigma$ for $z<0.042$, very close to the present epoch with a lower bound of $w_0=-0.9865$. However, the $\Lambda$CDM model remains included within 2$\sigma$ for the reconstructed CPL dark energy parametrization. Our result shows that the upcoming LSST supernovae data can help in restricting the parameter spaces of $\Lambda$CDM and CPLCDM. Thus, we hope that a similar analysis applied to more exotic dark energy models \citep{Zhai:2017vvt, Dhawan:2020xmp} may lead to better constraints in their respective parameter spaces, and hence imply interesting results. We leave the exploration of such for future work.

\section{Conclusions} \label{sec:conclusion}

As the nature of dark energy remains elusive with tensions coming into play, it is of prime importance to revisit the present status of cosmology in view of upcoming data. Rubin LSST is expected to represent the epitome of accuracy for Type Ia supernova (SNIa) data in the near future. In this work, we examine the impact of future LSST data on two cosmological models of interest, mentioned in Sec. \ref{sec:model}. Our findings indicate that LSST can considerably enhance the precision of background quantities for both the vanilla $\Lambda$CDM and CPL parametrization of dark energy leading to better constraints on the model parameters. We further observe that LSST will be capable of eliminating the existing correlations between $H_0-\sigma_{8,0}$ and $\Omega_{m0}-\sigma_{8,0}$. Moreover, these two models offer no notable improvements to the Hubble or clustering tension when subjected to LSST three-year simulated data.

Compared to present constraints, our analysis with simulated LSST data shows small shifts in the value of $\sigma_{8,0}$ depending upon the model, though such deviations are statistically insignificant so far. Nevertheless, this motivates further exploration of the clustering tension in the context of other models by utilising LSST. Our future work is geared towards such investigations. Moreover, we have used Pantheon as the current SNIa dataset for the present study, and leave the analysis with Pantheon+ \citep{pantheon_new} for subsequent work.

LSST promises to be the next step in precision cosmology. In the current study, increased levels of precision mostly translated to increased tensions due to insignificant mean shifts. Nonetheless, it remains to be seen how other models, which have shown promise in resolving certain tensions, behave when LSST is taken into account. As mentioned earlier, the current LSST data used is from a three-year SNIa simulation. However, a future $10$-year LSST dataset with a considerably larger set of data points than 3-year data and the inclusion of more accurate systematics would definitely be more insightful. Further, space-based near infrared (NIR) SNIa data from ROMAN \citep{2021arXiv211103081R}, when combined with LSST data, could help provide additional information on the suitability of different dark energy models. 

In a nutshell, the primary features of the present analysis are as follows:

\begin{itemize}
    \item LSST promises to provide improved constraints on the cosmological parameters of any reasonable late-time dark energy model, with the possibility to do the same for other classes of models as well.
    \item Future LSST data shows potential in breaking correlations between certain parameters. This can have very significant consequences in the study of the Hubble tension and the clustering tension in the future.
\end{itemize}

We believe the entire analysis gives an insight into the role of LSST in reinvestigating existing models and into the constraining power of the data from LSST compared to existing SNIa datasets like Pantheon.

\section*{Acknowledgements}
We express our sincere gratitude to Richard Kessler and Surhud More for their insightful comments and helpful discussions during the course of this research. We would also like to thank Arko Bhaumik for his constructive feedback on the manuscript. RS thanks ISI Kolkata for financial support through Senior Research Fellowship. PM thanks ISI Kolkata for financial support through Research Associateship. SP thanks the Department of Science and Technology, Govt. of India for partial support through Grant No. NMICPS/006/MD/2020-21. We acknowledge the computational facilities of ISI Kolkata. We would also like to express our gratitude to the LSST DESC for granting us permission to utilise their proprietary SNIa data.


\section*{Data Availability}
The Rubin DESC authority has granted permission for the use of the LSST Rubin SNIa data published in \citet{Mitra:2022ykq}. LSST Rubin SNIa dataset used in this study can be shared with individuals on reasonable request sent to Ayan Mitra.



\bibliographystyle{mnras}
\bibliography{mnras} 








\bsp	
\label{lastpage}
\end{document}